\documentclass[10pt]{iopart}

\usepackage{epsfig}
\usepackage{graphicx}
\usepackage{dcolumn}
\usepackage{iopams}

\expandafter\let\csname equation*\endcsname\relax 
\expandafter\let\csname endequation*\endcsname\relax 

\usepackage{amsmath}
\usepackage[english]{babel}
\usepackage[utf8]{inputenc}
\usepackage{hyperref}
\usepackage{bm}
\usepackage{ifpdf}

\begin{document}

\title[Mass-imbalanced Three-Body Systems in Two Dimensions]{Mass-imbalanced Three-Body Systems in Two Dimensions}

\author{F F Bellotti$^{1,2,3}$, T Frederico$^{1}$, M T Yamashita$^{4}$, D V Fedorov$^{3}$, A S Jensen$^{3}$ and N T Zinner$^{3}$}

\address{
  $^{1}$Instituto Tecnol\'ogico de Aeron\'autica, DCTA,  12.228-900 S\~ao Jos\'e dos Campos, SP, Brazil\\
  $^{2}$Instituto de Fomento e Coordena\c{c}{\~a}o Industrial, 12228-901, S{\~a}o Jos{\'e} dos Campos, SP, Brazil\\
  $^{3}$Department of Physics and Astronomy - Aarhus University, Ny Munkegade, bygn. 1520, DK-8000 \AA rhus C, Denmark\\
  $^{4}$Instituto de F\'\i sica Te\'orica, UNESP - Univ Estadual Paulista, C.P. 70532-2, CEP 01156-970, S\~ao Paulo, SP, Brazil\\
}
\date{\today }

\begin{abstract}
We consider three-body systems in two dimensions with zero-range interactions for general 
masses and interaction strengths. The momentum-space Schr{\"o}dinger equation is solved
numerically and in the Born-Oppenheimer (BO) approximation. The BO expression is derived
using separable potentials and yields a concise adiabatic potential between the two
heavy particles.  The BO potential is Coulomb-like and exponentially decreasing at small and large distances, respectively. 
While we find similar qualitative features to previous studies, we find important quantitative 
differences. Our results demonstrate that mass-imbalanced systems that are accessible in 
the field of ultracold atomic gases can have a rich three-body bound state spectrum in two
dimensional geometries. Small light-heavy mass ratios increase the number of bound states. For $^{87}$Rb-$^{87}$Rb-$^{6}$Li and $^{133}$Cs-$^{133}$Cs-$^{6}$Li we find respectively 3 and 4 bound states.
\end{abstract}
\pacs{03.65.Ge, 21.45.-v, 36.40.-c, 67.85.-d}
\maketitle

\section{Introduction}\label{sec1}
Manipulating quantum systems on large, small and intermediate 
scales is no longer a distant dream. 
Ultracold quantum gases experiments have realized Bose-Einstein 
condensates \cite{bloch2008}
and degenerate Fermi gases \cite{ketterle2008} and study truly 
remarkable many-body states, but it has also proven its
worth in exploring the physics of few-body systems \cite{ferlaino2010}. 
Moreover, these studies can be conducted in 
different geometries as two- and one-dimensional quantum systems 
are produced regularly in laboratories \cite{bloch2008}.

The so-called Efimov three-body bound states were predicted over four decades ago \cite{efi70}. 
They arise when a three-body system, composed of equal mass particles, 
has all its two-body subsystems at the threshold of binding. In this case, infinitely 
many bound states with energies geometrically separated are expected. 
These are the Efimov states and this effect is called the Efimov effect.
The effect is related to a long-range effective force and it can 
occur even when the individual two-body forces have zero range, as it was 
shown in an analytically solvable model by Fonseca, Redish and Shanley 
\cite{fonsecaNPA1979}. This is 
an example of how long-range forces can arise in the three-body problem 
in a way unpredictable by two-body intuition. In the early work of
Ref.~\cite{fonsecaNPA1979} the Efimov problem was 
handled through the adiabatic approximation, more precisely the Born-Oppenheimer 
(BO) approximation that is commonly used in molecular physics.
The BO approximation considers a system composed of two heavy and one light particle. 
The terms {\it heavy} and {\it light} have relative meaning: two particles are 
heavier than the third one. In this approximation the heavy particles move 
very slowly while the light particle orbits around them. In fact, it is enough 
to assume that the heavy particles kinetic energy is (much) smaller than the 
kinetic energy of the light particle.

These results relating the Efimov effect and the adiabatic 
approximation were obtained for three-dimensional (3D) systems. The 
dynamics and properties of quantum systems drastically change when the 
system is restricted to lower dimensions. For example, the scattering-length 
is not well defined for two-dimensional (2D) systems \cite{adh1995} and the 
kinetic energy operator gives a negative centrifugal barrier for 2D systems 
with zero total angular momentum while the centrifugal barrier is always 
non-negative for 3D systems. Furthermore, it was shown that any infinitesimal 
amount of attraction produces a bound state in 2D \cite{landau1977,simon1976,artem2011a,artem2011b}, 
while a finite amount of attraction is necessary for binding a 3D system.

Another important difference between 2D and 3D systems is the occurrence of 
the Efimov effect. The effect does not occur in 2D neither for equal mass 
three-body system \cite{tjo75} or unequal mass systems \cite{lim1980}. 
While the BO approximation was implemented in Ref.~\cite{lim1980} to look for the 
Efimov effect in 2D imbalanced mass three-body systems, the 
mass-dependence of such systems was not addressed.
The importance of the mass dependence in the 2D imbalanced mass three-body 
systems was stressed in Refs.~\cite{bellottiJoPB2011,bellottiPRA2012}, where an 
increasing number of bound states was found for a decreasing mass of one particle. 
This situation, where one particle is much lighter than the other ones, 
is suitably handled in the adiabatic approximation.

In this work we study the BO approximation of 2D three-body systems 
from the mass-dependence perspective in a systematic fashion. 
We consider a 2D three-body system 
with zero-range interactions for general masses and interaction strengths. 
The momentum-space Schr{\"o}dinger equation is solved numerically both 
in full generality and in the BO approximation. The BO expression is 
derived using separable potentials and yields a concise adiabatic 
potential between the two heavy particles in the heavy-heavy-light system 
when the light particle coordinate is integrated out. 
The adiabatic potential is mass-dependent and it reveals an
increasing number of bound states for the decreasing mass of the light 
particle. We find a transcendental equation that describes the 
adiabatic potential and an approximate analytic expression as well 
as the asymptotic form. Furthermore, we show that the 
approximate analytic form of the potential is very close to the 
full adiabatic potential.

Furthermore, we estimate the number of bound states for a heavy-heavy-light system 
as a function of the light-heavy mass ratio. Infinitely many bound states 
are expected as this ratio approaches zero. However, for each given mass 
configuration, we still have a finite number of bound states. Our results 
demonstrate that mass-imbalanced systems that are accessible in recent ultracold 
atomic gases experiments can have a rich three-body bound state spectrum 
in 2D geometries, where small mass ratios increase the number of bound 
states. For $^{87}$Rb-$^{87}$Rb-$^{6}$Li and $^{133}$Cs-$^{133}$Cs-$^{6}$Li 
we find respectively 3 and 4 bound states. 

The paper is structured as follows. The introduction in Section~\ref{sec1} 
is followed by the 2D 
three-body system formalism for zero-range interactions and general 
masses in the BO approximation in Section~\ref{sec2}.
The BO approximation and the adiabatic potential are shown 
in Section~\ref{sec3}. The results are presented in Section~\ref{sec4} and a discussion is 
given in Section~\ref{sec5}.    

\section{Formalism}\label{sec2}
\begin{figure}[!htb]
\centering
\includegraphics[width=0.5\textwidth]{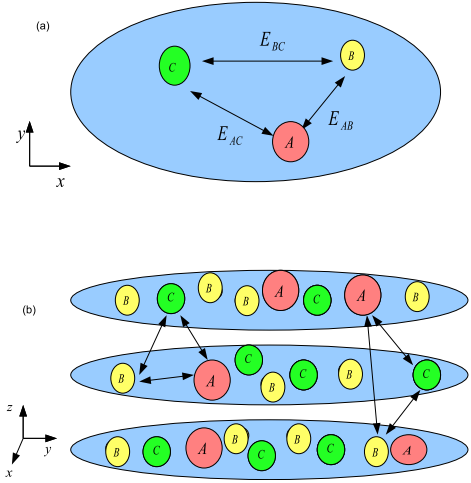}
\caption{A pictorial representation of the setup under consideration. 
(a) A three-body system in a plane (viewed from the top). We show the plane as a pancake with 
ellipsoidal shape rather than as a square depiction in order to convey the
fact that real system may have deformation of the trap that keep the particles
fixed in two dimensions. The three distinct particles are labeled $A$, $B$ and $C$. 
The two-body binding energies, $E_{AB}$, $E_{AC}$ and $E_{BC}$, and the 
coordinate axis are shown. (b) A multi-layer setup with
several $A$,$B$ and $C$ particles (side view). Long-range interactions 
in and between the layers can lead to bound state formation both within the 
same layer and for particles in different layers. In the case of three-body 
states, this is indicated by
black arrows.} 
\label{fig1}
\end{figure}

The general setup that we consider has particles that are always confined to a 
two-dimensional plane of motion, i.e. the kinematics is always 2D. However, 
we do not rule out systems with long-range interactions across many 
two-dimensional planes (as done in recent experiments with polar molecules \cite{miranda2011}).
Both of these situations are shown in figure~\ref{fig1}
for the most general case with three different particles $A$, $B$ and $C$.
The only assumptions concerning the interactions is that they are 
dependent only on the relative distance of the particles under 
consideration. We can then assume that 
the three-particle dynamics effectively happens in a single plane, 
since we may consider the layer index as an additional quantum 
number that specifies a few-body state.

Here we are considering only zero-range
interactions since we are interested in the model-independent 
low-energy universal limit. This limit emphasizes universal 
behaviour that should be independent of the particular 
system under study. In general, the validity of the 
approximation can be established by comparing the 
typical energy scale of the two-body interactions 
at short-range, i.e. for atomic systems this would 
typically be the van der Waals length, $r_\textrm{vdW}$.
In order for the bound states to be universal and 
the zero-range interaction limit to apply, we need the
binding energy, $|E_3|$, to satisfy 
$|E_3|\ll \tfrac{\hbar^2}{mr_\textrm{vdW}^{2}}$. We will 
assume that we are in this universal regime throughout
the paper. The zero-range limit provides a simplification of the 
Faddeev equation for the three-body bound state due to the 
separability of the zero-range interaction.
In the case where we have long-range interaction
across different layers and we have to assume that at low energy, the 
long-range forces may be approximated by a short-range pseudopotential 
that reproduces the two-body properties of the full potential. 
As discussed in Ref.~\cite{verhaar1985}, this can be done for 
potential that decay faster than $r^{-2}$ for $r\to\infty$. 
This includes the case of dipolar $r^{-3}$ interactions that 
we will return at the end of this paper.

The 2D Hamiltonian for the three-particle $ABC$ system with 
pairwise two-body potentials is
 \begin{equation} 
 H=H_0+V_{AB}+V_{AC}+V_{BC}
\end{equation}
where the three particles are, in principle, assumed to be bosons. However, 
for three different mass particles there is no symmetry requirements. Our
formalism takes the symmetry fully into account when two particles 
are identical bosons.
The kinetic energy
operator is
\begin{equation}
H_0=\frac{\mathbf{q}_A^2}{2\mu_{A,BC}}
+\frac{\mathbf{p}^2_A}{2\mu_{BC}}
=\frac{\mathbf{q}_B^2}{2\mu_{B,AC}}
+\frac{\mathbf{p}^2_B}{2\mu_{AC}}
  =\frac{\mathbf{q}_{C}^2}{2\mu_{C,AB}}
+\frac{\mathbf{p}^2_{C}}{2\mu_{AB}},
\end{equation}
where we use Jacobi relative momenta given in terms of
rest frame momenta,  $\mathbf{k}_i$ with $i=\alpha,\beta,\gamma$,
as
\begin{equation}
\mathbf{q}_\gamma=\mathbf{k}_\gamma \,\, \textrm{and} \, \, 
\mathbf{p}_\gamma=\mu_{\alpha\beta}\left(\frac{\mathbf{k}_\alpha}{m_\alpha}-\frac{\mathbf{k}_{\beta}}{m_\beta}\right)
\label{jacobi}
\end{equation}
where $(\alpha,\beta,\gamma)$ is the cyclic permutations of the
particles $A$, $B$, and $C$ with masses $m_A$, $m_B$ and $m_C$. The
reduced masses are $\mu_{\alpha\beta}=\frac{m_\alpha
m_\beta}{m_\alpha+m_\beta}$ and
$\mu_{\gamma,\alpha\beta}=\frac{m_\gamma(m_\alpha
+m_\beta)}{m_\alpha+m_\beta+m_\gamma}$. We use a separable potential with
operator form
\begin{equation}
V_{\alpha\beta}=\lambda_{\alpha\beta}|\chi_{\alpha\beta}\rangle\langle\chi_{\alpha\beta}|,
\end{equation}
where the form factor is $\left\langle \mathbf{p}_\gamma \right|\left.\chi_{\alpha\beta}\right\rangle \ = g(\mathbf{p}_\gamma)$,
depends only on the relative momentum of the two particles $\alpha$ and $\beta$. 
The limit to zero-range interaction is simply $g(\mathbf{p})=1$.

In order to relate the separable potential ansatz above to physical properties of 
the two-body system we use the the two-body T-matrix. For negative energies and zero-range
potentials it is defined by
\begin{equation}
T_{\alpha\beta}(E)=|\chi_{\alpha\beta}\rangle\tau_{\alpha\beta}(E)\langle\chi_{\alpha\beta}|,
\label{T}
\end{equation}
where, the matrix element of the 2D transition matrices are given by
(see e.g. Ref.~\cite{adh88} for the case of identical particles)
\begin{eqnarray}
\tau_{\alpha\beta}(E)=\left[ \frac{-4\pi \mu_{\alpha\beta}}{\hbar^2} \ln \left( \sqrt{\frac{E}{E_{\alpha\beta}}}\right)
\right] ^{-1} \ , \label{tau}
\end{eqnarray}
where $\alpha,\beta=A$, $B$ or $C$ and $\alpha \neq \beta$. $E_{\alpha\beta}$ 
is the energy of the $\alpha\beta$ two-body bound states. We will measure all three-body
energies in units of the two-body energy. In general systems where $A$, $B$, and
$C$ are different, one only needs to make a specific choice between the $E_{\alpha\beta}$. 
We will explicitly state our units in these cases. Also it is important to
know that $E_{\alpha\beta}=0$ it corresponds to a non-interacting $\alpha\beta$
system since $\tau_{\alpha\beta}\to 0$ in this case. This is different 
from 3D where a two-body bound state at zero energy happens at unitarity, 
i.e. where the interactions are as strong as allowed by the unitary 
limit of quantum mechanics applied to the $s$-wave scattering amplitude.

In order to solve for three-body bound states, the wave function $|\Psi_{ABC}\rangle$ 
is decomposed in
terms of the Faddeev components as
\begin{equation}
|\Psi_{ABC}\rangle=|\Psi_{A}\rangle+|\Psi_{B}\rangle+|\Psi_{C}\rangle
\end{equation}
where
$|\Psi_{\gamma}\rangle=G_0(E)V_{\alpha\beta}|\Psi_{ABC}\rangle$ with
the resolvent $G_0(E)=(E-H_0)^{-1}$ which is nonsingular for bound
states.
The bound state equation can be solved by the method described in
Ref.~\cite{bellottiJoPB2011} for $AAB$ systems with some straightforward
generalizations to the $ABC$ structure. We therefore refer to 
Ref.~\cite{bellottiJoPB2011} for the details of the formalism and do 
not repeat them here. Instead we now proceed to discuss the adiabatic
approximation.

\section{The Adiabatic Approximation}\label{sec3}
\begin{figure}[!htb]
\centering
\includegraphics[width=0.6\textwidth]{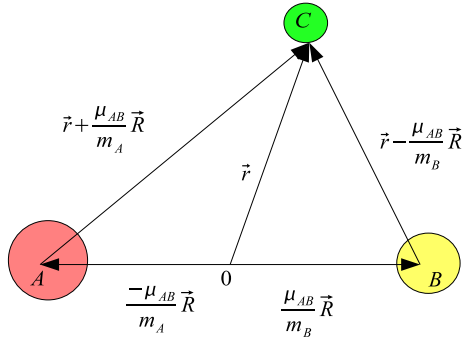}
\caption{Three-body relative coordinates used in the adiabatic approximation. Heavy 
particle of mass $m_A$ and $m_B$ interact with a light particle of mass $m_C$. The 
heavy particles are fixed at a distance $\bm R$ from each other and the origin of 
the coordinate system is chosen as the center-of-mass of the two heavy particles.
The coordinate of the light particle is its distance from this choice of origin.} 
\label{Graph01}
\end{figure}

We now consider two heavy particles with mass $m_A$ and $m_B$. 
These particles are fixed and their centers are separated by a 
distance $\mathbf{R}$. The additional light particle has mass $m_C$ and 
coordinate $\mathbf{r}$ relative to the center-of-mass of the heavy subsystem. 
The configuration and coordinates are 
depicted in figure~\ref{Graph01}.
We assume that the particles interact with each other through 
short-range potentials. The notation for the potential is that 
$v_C$ means the interaction between particles $A$ and $B$ and 
$v_A$ and $v_B$ are analogously defined. 
The three-body Hamiltonian is given by 
\begin{align}
H&=\frac{p_{C}^{2}}{2\mu_{AB}}+\frac{q_{C}^{2}}{2\mu_{C,AB}}+v_A+v_B+v_C \ ,& \label{eq.01}
\end{align}
where the Jacobi relative momenta and reduced mass definitions are given 
in (\ref{jacobi}). The Schr{\"o}dinger eigenvalue equation is $H\Psi(\mathbf{r,R})=E\Psi(\mathbf{r,R})$
where the Hamiltonian, $H$,
is given by
\begin{align}
&H=-\frac{\hbar^2}{2\mu_{AB}}\nabla^2_R-\frac{\hbar^2}{2\mu_{C,AB}}\nabla^2_r
+v_A\left(\mathbf{r}-\frac{\mu_{AB}}{m_B}\mathbf{R}\right)
+v_B\left(\mathbf{r}+\frac{\mu_{AB}}{m_A}\mathbf{R}\right)
+v_C(\mathbf{R}).& 
\label{eq.02}
\end{align}

The adiabatic approximation now instructs a split of the three-body eigenvalue 
equation into the solution of two two-body problems: the light particle motion 
is considered with respect to the heavy-heavy system and the heavy-heavy system 
motion is separated out. These eigenvalue equations will be useful whenever 
the motion of the light particle is rapid compared to the motion of the heavy 
ones, so that the light particle dynamics can be solved while the heavy 
particles are instantaneously at rest. The wave function is written as
\begin{equation}
\Psi(\mathbf{r,R})=\psi(\mathbf{r,R})\phi(\mathbf{R}) \ ,
\label{eq.03}
\end{equation}
where $\psi(\mathbf{r,R})$ is the wave function describing the motion of the light 
particle for fixed $\mathbf{R}$ and $\phi(\mathbf{R})$ is the 
heavy-heavy system wave function. The approximation is valid when the kinetic 
energy, $-\frac{\hbar^2}{2\mu_{AB}}\nabla^2_R \psi(\mathbf{r,R})$, is 
small compared to the other terms in (\ref{eq.02}). Using the wave 
function (\ref{eq.03}), we write the light particle eigenvalue equation as
\begin{equation}
\left[-\frac{\hbar^2}{2\mu_{C,AB}}\nabla^2_r+v_A\left(\mathbf{r}
-\frac{\mu_{AB}}{m_B}\mathbf{R}\right)+v_B\left(\mathbf{r}+
\frac{\mu_{AB}}{m_A}\mathbf{R}\right)\right]\psi(\mathbf{r,R})=\epsilon(R)\psi(\mathbf{r,R}).
\label{eq.05}
\end{equation}  
where $\epsilon(R)$ is a separation constant, which according to this 
equation cannot depend on $\mathbf{r}$. Furthermore, $\epsilon(R)$ plays 
the role of an effective potential in the equation describing the 
heavy-heavy particle system, i.e.
\begin{equation}
\left(-\frac{\hbar^2}{2\mu_{AB}}\nabla^2_R+v_C(\mathbf{R})+\epsilon(R)\right)\phi(\mathbf{R})=E\phi(\mathbf{R}) \ .
\label{eq.06}
\end{equation}
Assuming that the potentials in (\ref{eq.05}) are separable and 
have the same strength, i.e. 
$v_\alpha=\lambda \left\vert \chi_\alpha \right\rangle \left\langle \chi_\alpha \right\vert$, 
the light particle wave function in (\ref{eq.05}) in momentum space reads
\begin{equation}
\tilde{\psi}(\mathbf{p})=\lambda \frac{g(\mathbf{p})}{\epsilon(R)-\frac{\hbar^2}{2\mu_{C,AB}}p^2} \left[ e^{\imath \frac{\mu_{AB}}{m_A}\frac{\mathbf{p \cdot R}}{\hbar}} A_{+}+ e^{-\imath \frac{\mu_{AB}}{m_B}\frac{\mathbf{p \cdot R}}{\hbar}} A_{-} \right] \ , \label{eq.12}
\end{equation}
where
\begin{align}
A_+ &=\int d^{2}r^{\prime }\tilde{g}^{\dagger}\left(\mathbf{r^{\prime}} +\frac{\mu_{AB}}{m_A}\mathbf{R}\right) \Psi \left( \mathbf{r}^{\prime}\right)=\int d^{2}p^{\prime }g^{\dagger}\left(\mathbf{p^{\prime}}\right) \frac{e^{-\imath\frac{\mu_{AB}}{m_A} \frac{\mathbf{p^{\prime } \cdot R}}{\hbar}}}{2\pi\hbar}\tilde{\psi} \left( \mathbf{p}^{\prime}\right)  \ ,& \label{eq.11}\\
A_- &=\int d^{2}r^{\prime }\tilde{g}^{\dagger}\left(\mathbf{r^{\prime}} -\frac{\mu_{AB}}{m_B}\mathbf{R}\right) \Psi \left( \mathbf{r}^{\prime}\right)=\int d^{2}p^{\prime }g^{\dagger}\left(\mathbf{p^{\prime}}\right) \frac{e^{+ \imath \frac{\mu_{AB}}{m_B}\frac{\mathbf{p^{\prime } \cdot R}}{\hbar}}}{2\pi\hbar}\tilde{\psi} \left( \mathbf{p}^{\prime}\right)  \ .& \label{eq.11a}
\end{align}
Rewriting (\ref{eq.12}) in terms of $A_\pm$, we find the system of equations
\begin{equation}
A_\pm=\lambda\int d^2p \frac{\left\vert g(\mathbf{p})\right\vert^2}{\epsilon(R)-{\frac{\hbar^2}{2\mu_{C,AB}}p^2 }}\left( e^{\mp\imath \frac{\mathbf{p}\cdot \mathbf{R}}{\hbar}} A_\mp+ A_\pm \right). \label{eq.13}
\end{equation}
The non-trivial solution of (\ref{eq.13}), i.e $A_\pm \neq 0$, gives a transcendental equation for the energy, which reads
\begin{equation}
\frac{1}{\lambda}=\int d^2p \frac{|g(\mathbf{p})|^2}{\epsilon(R)-{\frac{\hbar^2}{2\mu_{C,AB}}p^2 }}\left[1+ \cos \left(\frac{\mathbf{p}\cdot \mathbf{R}}{\hbar}\right) \right]. \label{eq.15}
\end{equation}
We now use the binding energy of the two-particle heavy-light subsystem, $E_2$, and the T-matrix to eliminate $\lambda$ \cite{adh1995}. This yields
\begin{equation}
\int d^2p |g(\mathbf{p})|^2\left[\frac{1+  \cos \left(\frac{\mathbf{p}\cdot \mathbf{R}}{\hbar}\right)} {\epsilon(R)-\frac{\hbar^2}{2\mu_{C,AB}}p^2}+\frac{1}{|E_{2}|+\frac{\hbar^2}{2\mu_{C,AB}}p^2}\right]=0 \ . \label{eq.16}
\end{equation}

Since we are interested in low-energy, model-independent bound state, 
we describe the heavy-light particle system interacting through zero-range 
interactions. In momentum space this means that $g(\mathbf{p})=1$, so 
(\ref{eq.16}) is finite and its two terms can be integrated 
to give the transcendental equation 
\begin{align}
&\log \frac{\left\vert \epsilon(R)\right\vert }{\left\vert E_2 \right\vert} =2 K_0 \left(\sqrt{\frac{2\mu_{C,AB}|\epsilon(R)|}{\hbar^2}}R \right)  \ ,& \label{eq.51}
\end{align}
where $K_0$ is the zero order modified Bessel function of the second kind.
The (\ref{eq.51}) is a powerful tool in understanding mass-imbalanced 
three-body systems in two dimensions. We are able to solve it to get an 
analytic form for the small and large distance behavior of the adiabatic 
potential. We can also numerically solve it to test the validity of the 
analytical expressions.

When the separation, $\mathbf{R}$, between the two-heavy-particles is 
large, i.e. $\left\vert\mathbf{R}\right\vert \to \infty$, the light 
particle feels only the interaction from one of the heavy particles. 
In this limit the three-body problem becomes a two-body problem and we 
expect that $\left\vert E\right\vert =\left\vert E_2\right\vert$. 
Thus, defining 
$\left\vert \epsilon(R) \right\vert =\left\vert E_2\right\vert+V(R)$, 
this condition is fulfilled when $V\rightarrow 0$ for $\vert\mathbf{R}\vert\rightarrow \infty$. 
Inserting these definitions in (\ref{eq.51}) 
and expanding both sides up to first order in $V(R)$, we obtain
\begin{equation}
V(R)=\frac{2 \left\vert E_2\right\vert K_0 \left(s(R) \right)}{1+ s(R) \ K_1 \left(s(R)\right)} ,
\label{eq.35}
\end{equation}
where $s(R)=\sqrt{\frac{2\mu_{C,AB}\vert E_2\vert}{\hbar^2}} \ R$. In fact, 
(\ref{eq.35}) is a very good approximation of the BO potential $\epsilon(R)$.
We are also able to solve (\ref{eq.51}) to get 
the small distance behavior of the adiabatic potential. Using the asymptotic 
form of $K_0$ for small arguments \cite{abra}, the adiabatic potential reads
\begin{align}
\frac{|\epsilon_{asymp}(R)|}{|E_2|}\to
\frac{2 e^{-\gamma}}{s(R)} \left(1-\frac{e^{-\gamma}}{2} s(R)\left[(1-\gamma)- \frac{1}{2} \ln\left(\frac{e^{-\gamma}}{2}s(R)\right) \right]  \right)^{-1}\label{adpot1-subnum}
\end{align}
for $R\to 0$ (where $\gamma$ is Euler's constant). For large arguments, $R\to\infty$, the adiabatic 
potential becomes
\begin{align}
\frac{|\epsilon_{asymp}(R)|}{|E_2|}\to
1+ \frac{2 K_0 \left(s(R) \right)}{1+ s(R) \ K_1 \left(s(R)\right)}\,\textrm{for}\,R\to\infty.\label{adpot2-subnum}
\end{align}

Once the adiabatic potential is known, we can solve (\ref{eq.06}) 
and determine the bound state spectrum. In order to write 
a Sturm-Liouville eigenvalue equation for the heavy-heavy system in 
a $L_z=0$ state, we set $\phi =\frac{\chi}{\sqrt{R}}$ and (\ref{eq.06}) becomes
\begin{equation}
\left[-\frac{\hbar^2}{2\mu_{AB}}\left(\frac{\partial ^{2}}{\partial R^{2}} 
+\frac{1}{4R ^{2}}\right)+v_C(\mathbf{R}) +\epsilon \left( R \right) \right] \chi(R) = E \ \chi(R)
\label{eq.43}
\end{equation}
with the adiabatic potential from (\ref{eq.51}) given in the limits of
$R \to 0$ and $R \to \infty$ by (\ref{adpot1-subnum}) and
(\ref{adpot2-subnum}) respectively.

Due to the behavior of the adiabatic potential at small ($1/R$ Coulomb-like) and large distance
(exponential decrease with $\sqrt{m_C}$), 
we expect to find an increasing number of bound states when particle $C$ 
is much lighter than the other ones ($m_C \to 0$). In this limit the 
adiabatic potential becomes more attractive and these states will 
accumulate both at $R\to 0$ as 
the strength of the $1/R$ Coulomb-like potential increases, see
(\ref{adpot1-subnum}), and at $R\to \infty$, where more states are
allowed because the exponential cut-off moves to larger distances,
see (\ref{adpot2-subnum}).  However, for finite $m_C$, the number
of bound states is still finite. 

Although we can not address the limit $|E_2| \to 0$ in (\ref{adpot1-subnum}) 
and (\ref{adpot2-subnum}) in the current paper, 
it is worthwhile to point out that the limit where all subsystems interacting 
through zero-range interactions are unbound ($|E_2|\to 0$)
does not support three-body 
bound states in two dimensions \cite{tjo75,lim1980,bellottiPRA2012}. However, recent 
studies show that these so-called Borromean states are possible in 2D 
but only for potentials with an outer barrier and an inner attractive
pocket \cite{nie97,volosniev2012}.

\section{Results}\label{sec4}
\subsection{Adiabatic potentials}
We found analytical forms for the small, (\ref{adpot1-subnum}), and 
large, (\ref{adpot2-subnum}), distance behavior of the adiabatic 
potential. Once we are able to numerically compute the exact 
adiabatic potential from (\ref{eq.51}), 
we must show that our asymptotic behaviors in 
(\ref{adpot1-subnum}) and (\ref{adpot2-subnum}) are
consistent with the exact adiabatic potential.  

\begin{figure}[!htb]
\centering
\includegraphics[width=0.5\textwidth]{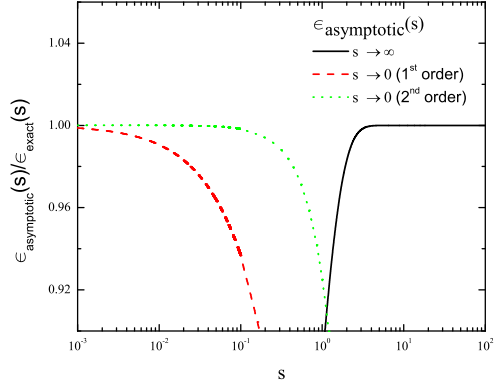}
\caption{Ratio $\epsilon_{asymptotic}(s)/\epsilon_{exact}(s)$ as
  function of the dimensionless coordinate $s$, showing the validity
  of the asymptotic behavior in (\ref{adpot1-subnum}) and
  (\ref{adpot2-subnum}).  The solid (black) and dashed (red) curves are
  the first order expansion at large and small distances,
  respectively.  The dotted (green) curve is the second order expansion
  (\ref{adpot1-subnum}).} 
\label{Graph03}
\end{figure}

The exact potential, $\epsilon_{exact}(s)$, obtained numerically from
(\ref{eq.51}) is shown in figure~\ref{Graph03} and compared to
$\epsilon_{asymp}(s)$ of equations.~(\ref{adpot1-subnum}) and
(\ref{adpot2-subnum}).  The asymptotic behavior does indeed work very
well for most of coordinate space. The largest deviations
are found in the region $0.3<s<3$, where the difference between
$\epsilon_{asymp}(s)$ and $\epsilon_{exact}(s)$ still never exceeds $9\%$.
Notice that the approximation accuracy increases a lot when higher
order terms are included in the expansions. One could go to more
precise adiabatic potential representations taken higher order
expansions of (\ref{eq.51}).  However, the results of the
approximate and full adiabatic potentials are almost 
indistinguishable in practice.

\subsection{Bound states}
It is known that the number of bound states increases as one
particle becomes much lighter than the other ones
\cite{bellottiJoPB2011,bellottiPRA2012}.  Here, we have calculated the
adiabatic potential asymptotic forms (\ref{adpot1-subnum}) and
(\ref{adpot2-subnum}) and found a differential equation (\ref{eq.43})
applicable to the case where two particles are much heavier than the
third. In the following, we specialize to identical heavy particles 
i.e., $m_A=m_B=M$ and $E_{AC}=E_{BC}=E_2$. 
From now on, we also adopt units such that $M=\hbar=|E_2|=1$ 
for simplicity.
The mass ratio between light and heavy particles is then denoted
$m=\frac{m_C}{M}$ as in \cite{bellottiJoPB2011}.  In this case, the
reduced mass $\mu_{C,AB}$ is written as
\begin{equation}
\mu_{C,AB}=\frac{2m}{m+2} \ \text{and} \ \mu_{C,AB} \to m \ \text{for} \ m \to 0 \ .
\label{eq.25}
\end{equation}
With these definitions and denoting the three-body energy by $E=E_3$, a useful adiabatic potential expression is given by
\begin{align}
\epsilon(R)\to -\frac{2 e^{-\gamma}}{\sqrt{\frac{4m}{m+2}} \ R} \left(1-\frac{e^{-\gamma}}{2} \sqrt{\frac{4m}{m+2}}  \ R\left[(1-\gamma)- \frac{1}{2} \ln\left(\frac{e^{-\gamma}}{2} \sqrt{\frac{4m}{m+2}} \ R\right) \right]  \right)^{-1}\label{adpot1a-subnum}
\end{align}
for $ \sqrt{\tfrac{4m}{m+2}}  \ R \leq 1.15$  and 
\begin{align}
\epsilon(R)\to -1- \sqrt{2 \pi } \frac{e^{-\sqrt{\frac{4m}{m+2}} R}}{\sqrt{\left(\sqrt{\frac{4m}{m+2}}\right)^{\frac{1}{2}}R}}
\,\textrm{for}\,\sqrt{\frac{4m}{m+2}} R \geq 1.15.\label{adpot2a-subnum}
\end{align}
This approximation is very accurate when
$2R \approx 1.15 \sqrt{(1+2/m)}$ where the largest deviation of $9\%$
is reached.  Proper bound three-body states are present when $E_3-E_2\leq0$,
or equivalently $|E_3|\geq|E_2|$. 

We numerically solve the differential equation (\ref{eq.43}) with the
adiabatic potential (\ref{adpot1a-subnum}) and (\ref{adpot2a-subnum})
to estimate the number of bound states ($N_B$) for a system with mass
ratio $m$ when the heavy particles do not interact with each
other. Letting $E_{AB}$ be the binding energy of the heavy-heavy system,
we can choose $v_C=0$ in (\ref{eq.43}) and get $E_{AB}=0$. If $v_C$
is attractive and able to support bound states, the three-body system
would effectively be reduced to the lightest particle moving around a
heavy-heavy dimer. The corresponding additional much deeper-lying
bound states are, however, not interesting in the present context. 
Such states are not universal three-body structures that we address in 
the current study, but should rather be regarded as bound two-body 
states dressed by an orbiting third particle.
The numerical solution is obtained by writing the 
eigenvalue equation (\ref{eq.43}) 
in matrix form. After discretisation of the operators and the 
wave function of the radial equation, we get a tridiagonal matrix 
which can be diagonalized to obtain the energies and the 
number of bound states. As always, one needs to introduce both
a short-range and a long-range cut-off to numerically solve
the differential equation in (\ref{eq.43}). But changing the 
cut-offs in a systematic manner (making the short-range cut-off
smaller and the long-range cut-off larger) we have checked that 
our results for the number of bound states are properly converged 
(we discuss this quantitatively below).

\begin{figure}[!htb]
\centering
\includegraphics[width=0.5\textwidth]{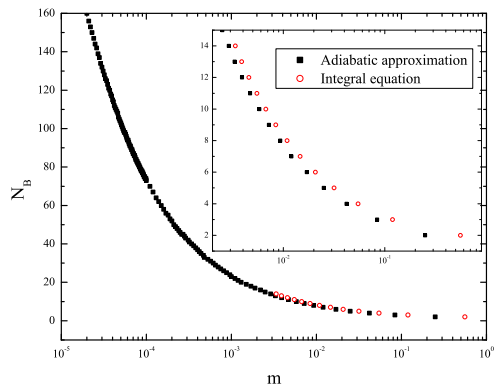}
\caption{Number of bound states, $N_B$, for a system with 
mass ratio $m$ and $E_{AB}=0$. The (black) squares represent the 
mass ratio $m$ where $N_B$ states are bound, calculated from
the adiabatic approximation in (\ref{eq.43}). The (red) circles represent 
the full numerical solution. The inset shows the result for the 
largest $m$ values on a different scale.}  
\label{Graph02}
\end{figure} 

The number of bound states as function of mass ratio, $m$, is shown
in figure~\ref{Graph02}, where a comparison 
between the adiabatic result of (\ref{eq.43}) and the full numerical solution is made.
Comparing both results in figure~\ref{Graph02}, we see that the adiabatic 
approximation picks up the small mass behavior very well, even for mass 
ratios up towards 1. There is a small error in the threshold for the 
number of available bound states for $0 \leq N_B \leq 14$,  but it 
decreases as $m \to 0$. The adiabatic approximation has an accuracy 
better than $10\%$ for $m=0.01$, as  can be seen in the inset of 
figure~\ref{Graph02}. Due to the numerical difficulties, it is very 
hard to count the number of bound states for $N_B>14$ by solving 
the full problem numerically. Fortunately, it is very easy to do it 
through the differential equation (\ref{eq.43}) with the adiabatic 
potentials (\ref{adpot1a-subnum}) and (\ref{adpot2a-subnum}). We also
see clearly in figure~\ref{Graph02} that $N_B \to \infty$ for $m \to 0$, 
as pointed out in \cite{bellottiJoPB2011}.

A fit to the results presented in figure~\ref{Graph02} shows that 
the dependence of the number of bound states, $N_B$, with the 
mass ratio, $m$, is rather well described by
\begin{equation}
N_B \approx \frac{0.731}{\sqrt{m}}.
\label{numberbs}
\end{equation}
This behavior can be explained by the old quantum theory 
(perhaps better known as the the semi-classical JWKB approximation of
Jeffreys, Wentzel, Kramers and Brillouin).
What we need to consider is the number of nodes in the 
wave function at zero energy, since this measures the number
of allowed bound states. 
Within the old semi-classical quantum theory, the usual way to 
estimate the number of bound states in an 
one-dimensional quantum problem is
\begin{equation}
\int{p \ dq}=N \pi \hbar.
\label{BS}
\end{equation} 
This is the JWKB estimate of the number of bound states in a
given potential.
Taking into account the effective potential in (\ref{eq.43}) 
and proper units, the number of bound states can be 
estimated from the formula
\begin{equation}
N=\frac{1}{\pi \sqrt{2 m}} \int_0^\infty{dx \ \sqrt{\frac{m}{2x^2}-V(x)}}=\frac{0.733}{\sqrt{m}} \ ,
\label{estimate}
\end{equation}
where $V(x)$ is the adiabatic potential (\ref{adpot1a-subnum}) 
and (\ref{adpot2a-subnum}) with $x=\sqrt{\frac{4m}{m+2}}R$. 
One may object that the integral in (\ref{estimate}) diverges 
in both limits and can not be performed. Introducing a lower and 
an upper cut-off in the integral, which are the same used in the 
numerical calculation ($10^{-2}$ and $10^5$ respectively), we 
get that $N=\frac{0.766}{\sqrt{m}}$. This result approaches the 
estimate in (\ref{numberbs}) once the diverging term in
(\ref{estimate}) becomes less important as $m$ becomes smaller. 
The integral in (\ref{estimate}) is $m$-independent for 
$m\leq 0.001$ with a $10^{-2}$ cut-off,  implying that the term 
$m/x^2$ is negligibly small itself. The apparent divergences are 
due to the semi-classical estimate, and accurately removed by 
a cut-off at both small and large $x$. The true quantum mechanical 
number of states can then be fully recovered.

\begin{figure}[!htb]
\centering
\includegraphics[width=0.5\textwidth]{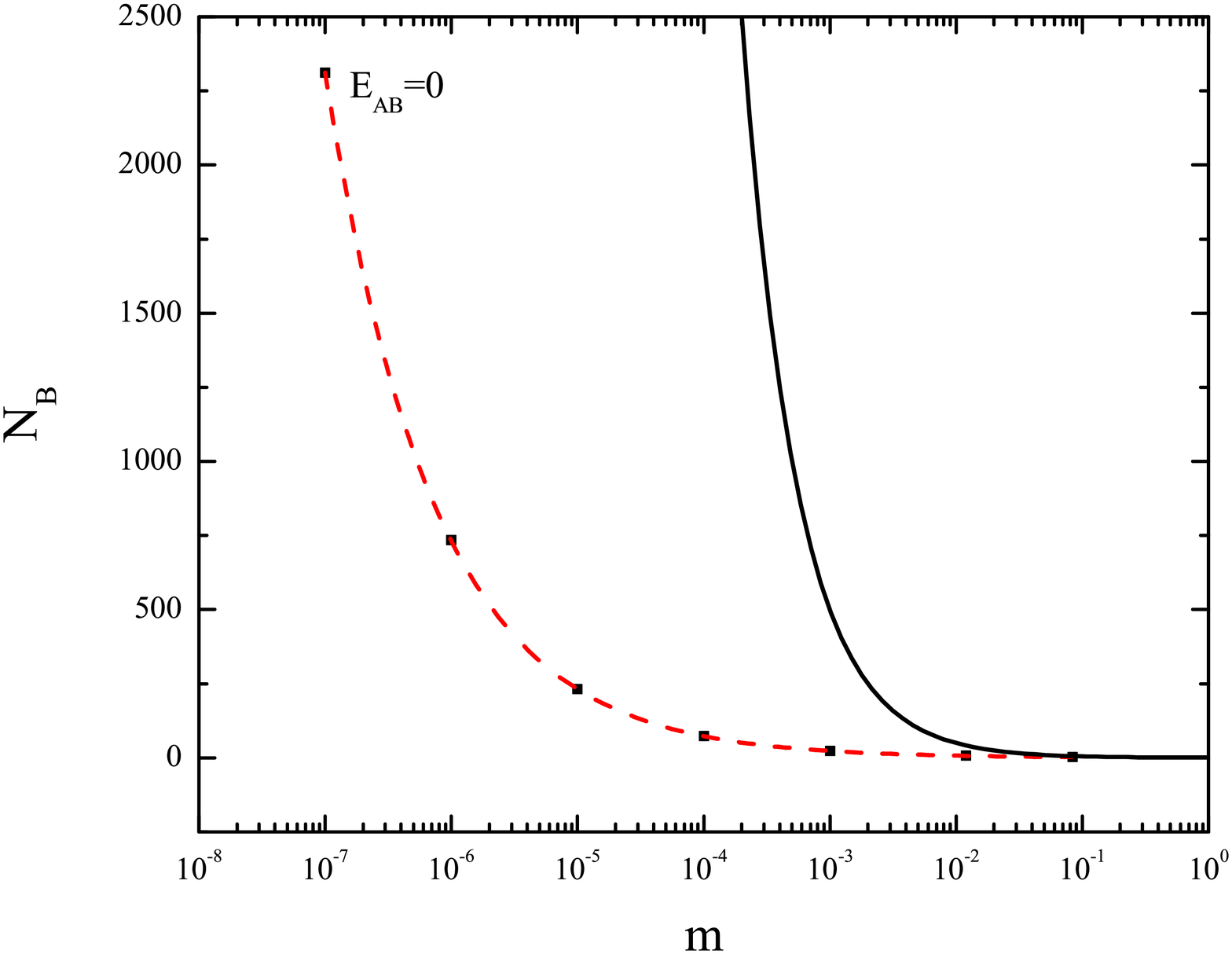}
\caption{Number of bound states, $N_B$, for a system with 
mass ratio $m$ for $E_{AB}=0$ (dashed (red) line) and for general 
(but fixed) $E_{AB}$ 
(full (black) line). The (black) squares represent some of the points 
from figure~\ref{Graph02}.} 
\label{Graph04}
\end{figure} 

The estimates of the number of bound states in (\ref{numberbs}) 
and (\ref{estimate}) agree very nicely. These estimates 
are less than the upper two-dimensional limit for a two-dimensional 
system with total angular momentum equal to zero, which is given 
in Ref.~\cite{math}. For the adiabatic potential in (\ref{adpot1a-subnum}) 
and (\ref{adpot2a-subnum}), this upper limit is given by 
$N=\frac{0.5}{m}$. The difference between the estimates is shown 
in figure~\ref{Graph04}. It is known that any three-body system in 
two dimensions will achieve its maximum number of bound states 
when all two-body subsystem are bound and have the same 
two-body binding energy \cite{bellottiPRA2012}. 
We thus expect that the estimate given by the dashed  curve 
in figure~\ref{Graph04} will hold for the adiabatic potential in
(\ref{adpot1a-subnum}) and (\ref{adpot2a-subnum}) when $E_{AB}=E_2$. 
Also, the number of bound states for a system with $0\leq |E_{12}| \leq 1$ 
is in the window between the solid curve and the dashed 
curve shown in figure~\ref{Graph04}. The large difference between 
these curves is necessarily due to states that have non-zero 
angular momentum that we are not concerned with in the 
present study but that are included in the strict mathematical 
bound cited above.

As we expected, the results confirm that the bound states accumulate
at both $R \to 0$ and $R \to \infty$ as $m \to 0$. From our numerical 
calculations we find that the energy of the
lower states (which reside at small distance) 
seem to increase without bound when $m\to 0$. Furthermore, in the 
same limit the wave function
vanishes slower at large distances, allowing more
bound states also at large distance. 
This can be interpreted as an Efimov-like effect for the
two dimensional case. However, an important distinction between the 2D
and 3D case must be made. While the Efimov effect says that three
identical particles can have infinitely many bound states when $E_2
\to 0$, in 2D this limit leads to an unbound three-body system. We
expected to have infinitely many bound states in 2D, when $m=0$ and
two interactive interactions. But we must stress that for finite 
$m$ we still have a finite number of bound states.

\subsection{Experimentally relevant systems}
Experiments with ultracold atoms are able to produce quasi-2D samples 
of $^{23}$Na \cite{sodium2D}, $^{40}$K \cite{modugno2003,gunter2005}, $^{87}$Rb 
\cite{rubidium2D}, $^{133}$Cs \cite{cesium2D1,cesium2D2,cesium2D3}, 
and $^{6}$Li \cite{sommer2012}.
It was recently reported that mixtures of $^{133}$Cs and $^{6}$Li were 
successfully trapped and that very favourable Feshbach resonances that 
can be used for tuning the 
interaction strength between the atoms have now been found \cite{repp2012,tung2012}. 
We note that the zero-range model of interactions that we use in the current 
study is applicable for broad Feshbach resonances where finite-range corrections
are negligible. Many heteronuclear systems typically have narrow Feshbach 
resonances where finite-range corrections are expected to play a larger
role \cite{chin2010}. However, the recent results of references~\cite{repp2012}
and \cite{tung2012} demonstrate that broad resonances are also available for
highly mass-imbalanced cases. We will therefore not consider finite-range
corrections and narrow resonances in this work.

In order to discuss some properties of mass-imbalanced 
three-body systems, we consider two different systems that can be
probed in laboratories in the near future. These system are $^{133}$Cs-$^{133}$Cs-$^{6}$Li, 
represented as circles in figures~\ref{Graph06} and \ref{Graph07}, 
and $^{87}$Rb-$^{87}$Rb-$^{6}$Li, represented as diamonds in the same figures.

\begin{figure}[!htb]
\centering
\includegraphics[width=0.9\textwidth]{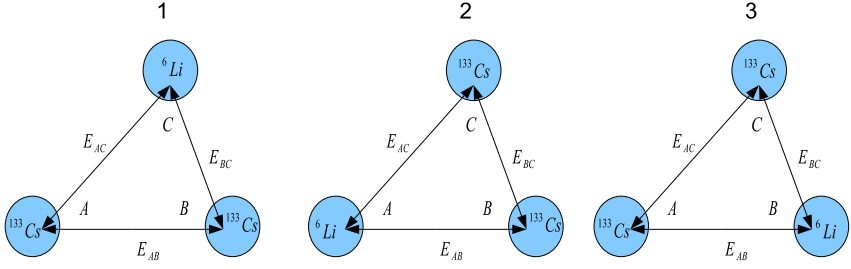}
\caption{Pictorials of the relation between masses and energies for regions $1$, $2$ and $3$ of 
shown in figures~\ref{Graph06} and \ref{Graph07}.} 
\label{Graph09}
\end{figure} 

Figures~\ref{Graph06} and \ref{Graph07} are mass-diagrams that show 
how many bound states a system 
composed of different particles supports for a given set of two-body 
binding energies in 
each subsystem. The energy symmetric case (all two-body binding energies
equal) was discussed in Ref.~\cite{bellottiPRA2012}. 
In general there are several ways to obtain a 
$^{133}$Cs-$^{133}$Cs-$^{6}$Li system if we assume that the interactions
in the different subsystems give rise to different binding energies. The
possiblities are shown in figure~\ref{Graph09}. If we consider a 
system of just one hyperfine state of $^6$Li and of $^{133}$Cs clearly 
two of the interactions will be identical (for instance $E_{AC}=E_{BC}$ on 
the left side of figure~\ref{Graph09}). However, as has been discussed 
in the recent studies of Feshbach resonances in this mixed system \cite{repp2012,tung2012}
different hyperfine states will in general give different interactions.
This implies that with different hyperfine states we may be able to 
access the full range of parameter space.
The adiabatic problem was handled for a non-interacting heavy-heavy 
system, i.e. $E_{AB}=0$. This is close to the situation in 
$^{133}$Cs-$^{6}$Li experiments, where three-body bound states are 
expected to be found when the subsystem $^{133}$Cs-$^{133}$Cs is 
almost non-interacting \cite{repp2012,tung2012}.
The relevant case is then $E_{AB}=0$ which is shown in figure~\ref{Graph06}.

\begin{figure}[!htb]
\centering
\includegraphics[width=0.5\textwidth]{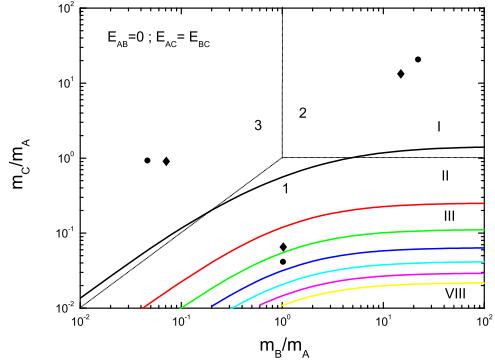}
\caption{Mass-diagram defining the mass ratios thresholds for 
the existence of $N$ bound states for $E_{AB}=0,E_{AC}=E_{BC}$.
The numbers $1$, $2$ and $3$ label three different regions according to 
figure~\ref{Graph09} and the Roman numerals denote the number 
of bound states present bellow each solid curve. The circles 
mark the $^{133}$Cs-$^{133}$Cs-$^{6}$Li systems and
the diamonds mark the $^{87}$Rb-$^{87}$Rb-$^{6}$Li systems.}
\label{Graph06}
\end{figure} 

The mass-diagram in figures~\ref{Graph06} and \ref{Graph07} was 
obtained by solving the full numerical integral equations as 
in Ref.~\cite{bellottiJoPB2011}. In figure~\ref{Graph06}, the 
points at which the lines cross $\tfrac{m_B}{m_A}=1$ corresponds to
the circles in figure~\ref{Graph02}. 
Notice that on region $1$, where 
$m_C<m_A,m_B$, the particle $C$ is always the \textit{lightest} 
and $E_{AB}$ is always the heavy-heavy particle energy. 
This region gives the deepest bound states 
in the adiabatic approximation. We see that the symmetry present in 
Ref.~\cite{bellottiPRA2012} is partially broken since $E_{AB}=0$
and only regions $2$ and $3$ are symmetric. 
The systems $^{87}$Rb-$^{87}$Rb-$^{6}$Li (diamond)
and $^{133}$Cs-$^{133}$Cs-$^{6}$Li (circles) 
have three and four bound states, respectively. This is the 
experimentally relevant case \cite{repp2012,tung2012}, and our results 
demonstrates that also in a 2D setup, the $^6$Li-$^{133}$Cs 
system will have several bound states.

It is known that the most favorable scenario for a 
spectrum with many bound states for two dimensional 
three-body systems is the symmetric 
energy case, where all the subsystems are equally bound, i.e. 
$E_{AB}=E_{AC}=E_{BC}$ \cite{bellottiPRA2012}. However, this 
scenario seems less likely in current experiments.
A very promising scenario is shown in figure~\ref{Graph06}, where 
the heavy-heavy system does not interact and the two heavy-light 
systems have the same binding energy. Except for the energy-symmetric 
case in Ref.~\cite{bellottiPRA2012}, this is the most favorable 
scenario for a rich energy spectrum in 2D.

Figure~\ref{Graph07} shows scenario where the two-body energies
are fully imbalanced. We have selected a generic case with 
energy scales related through $E_{AB}=10E_{AC}$ and $E_{BC}=0.1E_{AC}$. 
This may appear difficult
to set up experimentally, but fortunately this scenario does not seem 
to have any advantage over the others.
In region $1$, 
where our reference system has the heavy-heavy system more strongly 
bound than the others, the $^{87}$Rb-$^{87}$Rb-$^{6}$Li 
and $^{133}$Cs-$^{133}$Cs-$^{6}$Li systems have only one bound 
state each. In this energy configuration, the region $2$ should 
be the most similar to region $1$ in figure~\ref{Graph06}, 
where the heavy-heavy 
system is not as bound as the others. However, regions $2$ 
and $3$ appear almost symmetric in figure~\ref{Graph07}, showing 
that both systems have two bound states each. Effectively, we 
see that the strongly bound heavy-light system changes the threshold 
of binding for the three-body system and effectively removes 
the most weakly bound three-body states.

\begin{figure}[!htb]
\centering
\includegraphics[width=0.5\textwidth]{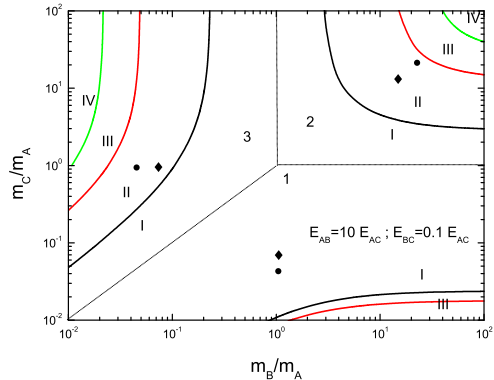}
\caption{Mass-diagram for an imbalanced energy system, defining 
the mass ratios thresholds for the existence of $N$ bound 
states for $E_{AB}=10E_{AC}$ and $E_{BC}=0.1E_{AC}$.
The numbers $1$, $2$ and $3$ 
label three different regions according to figure~\ref{Graph09} 
and the Roman numerals point out the number of bound states 
inside each solid curve.} 
\label{Graph07}
\end{figure}

\section{Summary and conclusion}\label{sec5}
We consider three-body systems with zero-range interactions for 
general masses and interaction strength in two-dimensions (2D). 
The aim of this work is to study the limit where one particle is 
much lighter than the others ($m_C \ll m_A=m_B$) for 
non-interacting heavy particles, i.e. $E_{AB}=0$. This limit contains
a rich energy spectrum and can be handled through the 
Born-Oppenheimer (BO) approximation. The BO approximation allows 
us to integrate out the light particle coordinate in order to 
derive the adiabatic potential between the heavy particles in 
the heavy-heavy-light system.

The adiabatic potential is found to be the solution of a transcendental 
equation. We are able to find analytical forms for the asymptotic behavior 
for large and small $R$, where $R$ is the distance between the two heavy 
particles. Comparing both asymptotic behavior and numerical solution 
of the adiabatic potential, we find that the analytic form precisely 
describes the adiabatic potential for any $R$ and can be directly
applied in 2D three-body system calculations. 
The analytic adiabatic potential explicitly shows the
mass-dependence. The analytic adiabatic potential 
becomes more attractive and suffers less screening as the mass of the light 
particle $C$ decreases ($m_C\to 0$). This potential explain the 
increasing number of bound states in this limit and we are 
able to demonstrate that the 
bound states accumulate both at $R\to 0$ and $R\to \infty$. 
We thus see an Efimov-like effect when $m_C=0$. However, 
we still caution that for any {\it finite} $m_C$ there are 
only a finite number of bound three-body states.

The results obtained with the analytic adiabatic potential agree 
with numerical solutions of the momentum-space Schr{\"o}dinger equation.
However, approaching the three-body problem in 2D with the 
analytic form of the adiabatic potential has the huge advantage that 
it is much less computationally demanding and allows for easy 
access to rich regions of the mass-diagrams. Furthermore, we 
are able to show that the number of bound states calculated 
with the analytic adiabatic potential matches the estimate 
given by the semi-classical JWKB methods of the old quantum theory, 
i.e. the Bohr-Sommerfeld method.

In recent work it was shown that the richest 2D energy spectrum is found when the 
three-body mass-imbalanced system has all two-body subsystems bound 
with the same energy, namely 
$E_{AB}=E_{AC}=E_{BC}$ for $m_C \ll m_A \neq m_B$ \cite{bellottiPRA2012}. 
However, as these conditions are difficult to obtain in experiment, 
we search for other configurations able to produce a rich energy spectrum. 
We numerically solve the momentum-space Schr{\"o}dinger equation for 
general masses and interaction strengths in order to build mass-mass 
diagrams, which outline the expected number of bound state for a general 
three-body system. We find that systems with equal mass and non-interacting 
heavy particles, i.e. $m_C \ll m_A=m_B$ and $E_{AB}=0;E_{AC}=E_{BC}$, 
are the most promising candidates to achieve a rich three-body 
bound state spectrum in 2D. Incidentally, 
this promising configuration is the one that
can be handled with great precision in the BO approximation as we have shown.

One of these promising configurations was recently reported as the 
experimental realization of $^{133}$Cs-$^{6}$Li systems \cite{repp2012,tung2012}. 
It even looks as if three-body
bound states can be expected when the subsystem $^{133}$Cs-$^{133}$Cs 
is almost non-interacting. We can estimate the number of bound states for 
such a system in a two-dimensional setup through the mass-mass diagrams in this work. 
For $^{87}$Rb-$^{87}$Rb-$^{6}$Li and $^{133}$Cs-$^{133}$Cs-$^{6}$Li we 
find respectively 3 and 4 bound states. It is very important to note 
that these numbers do {\it not} depend on the exact two-body energy in the 
$^6$Li-$^{133}$Cs subsystem. This two-body energy in the 2D setup are
functions of the 3D low-energy scattering length \cite{bloch2008} of 
the particular Feshbach resonance that is used in experiment to tune 
the interaction. However, as long as there is such a resonance, 
our results should apply when the system is squeezed into a 
two-dimensional geometry.

For now, the system $^{133}$Cs-$^{133}$Cs-$^{6}$Li prepared as in 
region 1 of figures~\ref{Graph09} and \ref{Graph06} seems to be the 
most promising realistic combination to achieve a richer three-body 
energy spectrum in 2D. Some other mixtures under intense study at 
the moment with large mass imbalances are Lithium-Ytterbium \cite{hansen2012} 
and Helium-Rubidium \cite{knoop2012}. Of particular interest is that 
there are different isotopes available in those systems so that the 
quantum statistics of the components can also be changed. In the
case of layered systems with long-range dipolar interactions as
shown in figure~\ref{fig1} is also extremely promising for 
finding bound 
states \cite{wang2006,jeremy2010,klawunn2010,baranov2011,jeremy2011,artem2012,zinner2012}, 
and we expect that some of these will have a universal low-energy 
character.
The possibility of tuning the 
binding energy of each pair and performing experiments mixing molecules 
and atoms should open new avenues for even richer two-dimensional 
three-body spectra. Another interesting discussion is what the 
trace of the three-body parameter is in the corresponding 
2D bound states. The three-body parameter
determines the overall scale of the three-body spectrum in 
3D and has recently been shown to be universally connected 
to the two-body van der Waals interaction 
\cite{berninger2011,naidon2011,chin2011,wang2012,schmidt2012,peder2012,knoop2012b}.
The 2D three-body problem does not require a three-body
parameter in order to study the universal limit with 
zero-range interaction due to the ever present two-body 
bound state with energy $E_2$ which makes the three-body
problem regular and well-defined even in the zero-range
limit \cite{tjo75}.
However, the fact remains 
that short-range physics due to the real atomic
two-body potential (typically of Lennard-Jones type) is
still present in the system. This clear physical 
origin of the three-body parameter 
indicates that it could still play a role in 2D in 
not only determining the precise physical value of 
$E_2$ for a given system but perhaps also have a 
quantitative influence on three-body energies.

In order to experimentally observe the presence of these three-body 
bound states one should be able to use similar techniques to those
used for the study of Efimov states in 3D, i.e. loss measurements
\cite{kraemer2006,ferlaino2010} and photo-association \cite{lompe2010,nakajima2011}.
It may also be possible to use RF (radio-frequency) transition techniques as in recent experiments
studied two-body bound states and many-body pairing in two-dimensional 
Fermi gases \cite{frohlich2011,sommer2012}. 
Another possible experimental signature of 2D three-body systems is
through the momentum distribution and the two- and three-body 
contact parameters which appears as 
coefficients \cite{werner2012a,bellotti2012,werner2012b}. 
These coefficients depends sensitively on the presence of 
bound two- and three-body states.

\paragraph*{Acknowledgments} This work was partly support by funds
provided by FAPESP (Funda\c c\~ao de Amparo \`a Pesquisa do Estado
de S\~ao Paulo) and CNPq (Conselho Nacional de Desenvolvimento
Cient\'\i fico e Tecnol\'ogico ) of Brazil, and by the Danish 
Agency for Science, Technology, and Innovation.

\end{document}